# The temperature-dependent magnetization profile across an epitaxial bilayer of ferromagnetic $La_{2/3}Ca_{1/3}MnO_3$ and superconducting $YBa_2Cu_3O_{7-\delta}$

**S. Brück[1,2], S. Treiber[2], S. Macke[2], P. Audehm[2], G. Christiani[3], S. Soltan[3], H.-U. Habermeier[3], E. Goering[2] and J. Albrecht[4]**

[1]University of Würzburg, Physikalisches Institut, Am Hubland, D-97074 Würzburg, Germany.

[2]Max Planck Institute for Metals Research, Heisenbergstr. 3, D-70569 Stuttgart, Germany.

[3]Max Planck Institute for Solid State Research, Heisenbergstr. 1, D-70569 Stuttgart, Germany.

[4]Hochschule Aalen, Beethovenstr. 1, D-73430 Aalen, Germany.

E-mail: seb@brueck-online.com

**Abstract**. Epitaxial bilayers of ferromagnetic $La_{2/3}Ca_{1/3}MnO_3$ (LCMO) and superconducting $YBa_2Cu_3O_{7-\delta}$ (YBCO) have been grown on single-crystalline $SrTiO_3$ (STO) substrates by pulsed laser deposition. The Manganese magnetization profile across the FM layer has been determined with high spatial resolution at low temperatures by X-ray resonant magnetic reflectivity (XRMR). It is found that not only the adjacent superconductor but also the substrate underneath influences the magnetization of the LCMO film at the interfaces at low temperatures. Both effects can be investigated individually by XRMR.

## 1. Introduction

Properties of heterostructures made of high-temperature superconductors such as YBCO and spin-polarized ferromagnets such as LCMO are strongly influenced by coupling phenomena at the interfaces. In general, superconductivity and ferromagnetism are incompatible ordering principles and their combination leads to substantial competition between the two ordered ground states [1]. In the case of hybrids of high-temperature superconductors and oxide ferromagnets, this usually leads to the suppression of both transition temperatures [2,3]. Several mechanisms have been discussed to explain this observation, including oxygen diffusion [4], charge transfer [5] due to a discontinuity in the chemical potential, a redistribution of the orbital occupancy [6], diffusion of spin-polarized quasiparticles [2,7], and finally dipolar magnetic coupling via stray fields [8,9]. In addition to lowered ordering temperatures, magnetic coupling phenomena can be observed in particular, in superlattices. Periodic structures of superconductors and ferromagnets with many layers, magnetic coupling between second nearest neighbours can be found indicating magnetic coupling via a superconducting layer [10]. This, however, could be experimentally shown only in superlattices where the superconducting part consists of $Y_{0.6}Pr_{0.4}Ba_2Cu_3O_{7-x}$ rather than $YBa_2Cu_3O_{7-x}$.

To obtain fundamental understanding of possible magnetic coupling phenomena in hybrid or heterostructures of YBCO and LCMO we focus in this paper on the simplest possible structure, a bilayer of LCMO and YBCO on a single-crystalline substrate. This system was investigated by XRMR, a technique which provides structural and element selective magnetic depth profiles with very high spatial resolution [11-13].

Determining the magnetization profile across the ferromagnetic layer by XRMR we can directly investigate the interaction at the LCMO/YBCO and the LCMO/substrate interface. Both results can act as a prerequisite for the understanding of coupling phenomena in LCMO/YBCO superlattices.

## 2. Experimental

A 5x5 mm² $SrTiO_3$ (STO) single crystal with (100) cut and polished surface is used as a substrate. The substrate has been chemically treated [14] to provide a $TiO_2$ terminated (100) surface; it was then mounted in a growth chamber with a base pressure of $5 \cdot 10^{-4}$ Pa. Bilayers of LCMO and YBCO with a nominal thickness of 400 Å each have been deposited by PLD at a growth temperature of 730° C at an oxygen pressure of 50 Pa. Full oxygenation has been achieved by annealing the films at 530° C in an oxygen atmosphere of $10^5$ Pa for ½ hour and then slowly cooling down to room temperature. These parameters lead to epitaxial growth of LCMO on STO and the resulting LCMO single layer film with a nominal thickness of 400 Å is a ferromagnetic metal with a $T^{Curie}$ of 240 K. Subsequently a 400 Å layer of optimally doped YBCO was grown on top of the LCMO. The superconducting transition temperature was experimentally derived from an identical sample using SQUID magnetometry and a value of $T_C$=88 K is found which verifies the high quality and optimal doping of the layer.

The XRMR and XMCD measurements were carried out at the undulator beamline UE56/2-PGM1 at BESSY II in Berlin, Germany. We use a dedicated soft X-ray magnetic reflectometry endstation built at the MPI for Metals Research [15]. This setup allows measuring XRMR curves for angles of incidence of $\Theta$=0° to 70°. Additionally high-quality X-ray absorption (XAS) and X-ray magnetic circular dichroism (XMCD) can be measured by total electron yield (TEY) and fluorescence yield (TFY) providing in-situ spectroscopy data necessary for reflectometry simulations.

In general it is not possible to obtain the optical profile directly from a measured reflectivity curve. Instead, an iterative approach is used to indirectly obtain the optical profile by fitting a simulated curve to the measurement. Using the simulation software ReMagX, which has been developed specially for this purpose at the MPI for Metals Research [16], we can simulate magnetic reflectivity and asymmetry curves and fit them to the measured reflectivity data. This task is further complicated by the fact that XRMR has to be measured at a magnetic active resonant absorption edge, i.e. a transition which shows circular dichroism (usually L/M edges in the soft X-ray regime). The optical properties at resonant edges show dramatic changes with the energy and the electronic configuration of the elements, e.g. oxidation state [17]. It is therefore imperative to experimentally obtain the optical and magneto-optical constants of the investigated sample [18]. By measuring the near-edge X-ray absorption fine structure (NEXAFS) and also the XMCD for the relevant energy range, the optical and magneto-optical constants can be derived (see e.g. in [19,20]). This information is then used for the simulation of the XRMR curves while the structural (roughness/thickness) parameters are free parameters for the fitting process. This method ultimately provides element selective structural and magnetic depth profiles of a layered sample with very high spatial resolution [11-13].

## 3. Results

### 3.1 Magnetometry & Magneto-Optics

SQUID magnetometry has been used to characterize the bulk magnetic properties of the LCMO/YBCO bilayer at T=100 K, i.e. above the superconducting transition. To avoid damaging the surface of the sample intended for the XRMR investigation, a sister sample simultaneously prepared in the same run for the deposition has been used for this measurement. The sample was mounted in a Quantum Design MPMS-XL 7T SQUID, cooled from room temperature to T=100 K in zero field, and a magnetic hysteresis was recorded with the magnetic field oriented parallel to the sample surface, i.e. in-plane geometry. Figure 1 shows the resulting T=100 K hysteresis loop of the sample which has been corrected for the diamagnetic contribution from the STO substrate.

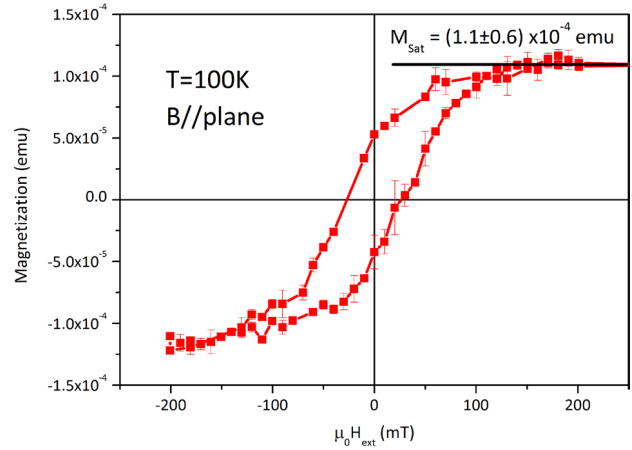

**Figure 1.** Magnetic in-plane hysteresis measured at T=100K using a SQUID magnetometer.

A clear ferromagnetic hysteresis with a coercive field of $H_C$=25 mT and a saturation magnetization of $M_{Sat}$=(1.1±0.6)·$10^{-4}$ emu is found. Using a value of 390 Å for the thickness of the LCMO layer as revealed by reflectometry (see below) and taking the lateral film size of 5x5 mm² (substrate size) into account, the saturation magnetization is $M_{Sat} = 113$ emu/cm³ which corresponds to a magnetic moment per Mn atom of $0.7\mu_B$. This value is much smaller compared to the bulk magnetization value of $3.7\mu_B$ for $La_{2/3}Ca_{1/3}MnO_3$ [21]. From previous investigations of LCMO/YBCO bilayers it is known that the magnetic moment of LCMO is affected by the presence and thickness of the YBCO layer [2]. The maximum saturation magnetization of LCMO decreases for increasing YBCO layer thickness and our results are comparable to those of Ref [2].

The superconducting properties of the YBCO layer were characterized using the magneto-optical Faraday Effect. This method uses an iron garnet indicator film to image the flux line density in the YBCO layer and allows determining the critical current density of the film [22]. Figure 2 shows images of the magnetic flux density distribution (left) and critical current density distribution (right) in the YBCO layer at a temperature of T=8 K measured in remanence [23]. The critical current density distribution was from the magnetic flux density distribution derived using the above described technique.

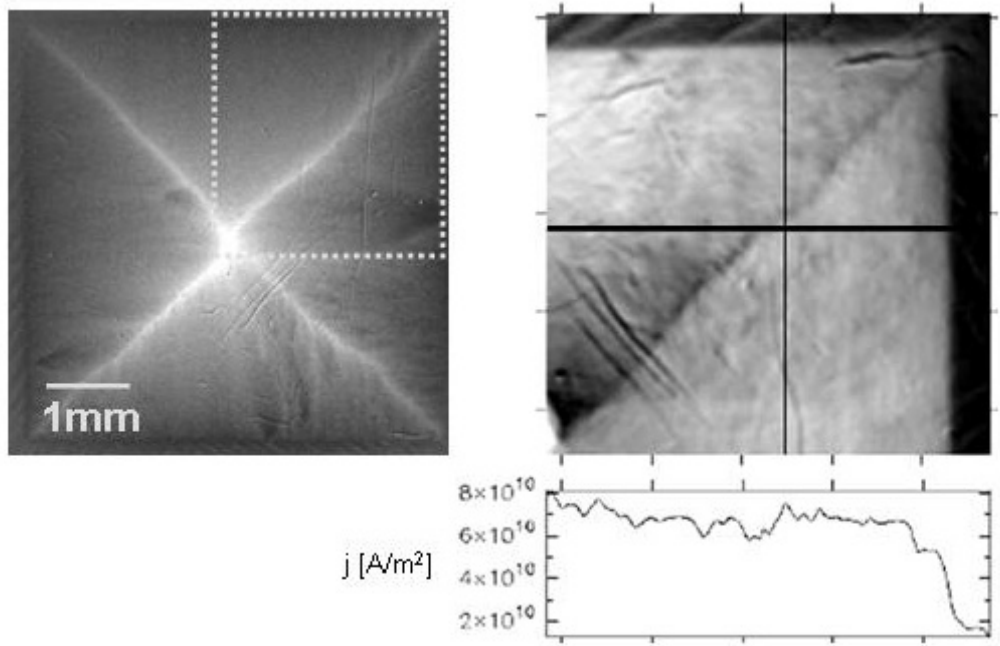

**Figure 2.** Magnetic flux density distribution (left) and calculated current density distribution (right) in the YBCO layer measured at T=8K in remanence. The average critical current density in the film is $7 \cdot 10^{10}$ A/m² as derived from the horizontal line profile (bold black line in the right image).

A maximal critical current density of $j_s = 7 \cdot 10^{10}$ A/m² is found. Compared to bulk YBCO, this value is reduced by nearly one order of magnitude which is related to the presence of the ferromagnet [24].

3.2 Optical Constants

The energy dependent optical properties of the individual layers, i.e. the complex indices of refraction $N_j(E) = 1 - \delta(E) + i\beta(E)$ where j is the layer index, are a necessary prerequisite for the determination of the structural and magnetic depth profile. While the imaginary (absorptive) part, β, is directly accessible from a soft x-ray absorption measurement by fitting the data to tabulated values [25], the real (dispersive) part, δ, can be calculated from β using the Kramers-Kronig relation [19].

An X-ray absorption spectrum (XAS) was measured by total electron yield (TEY) [26] in the energy range E=600 eV to E=700 eV. Linear polarized X-rays were used for the measurement and the sample was set to an angle of incidence of Θ=45°. The resulting spectrum was then adapted to tabulated X-ray scattering factors [25] to obtain the imaginary part of the scattering factor $f_2$ including the sample specific $L_{2,3}$ edge resonances. A Kramers-Kronig transform similar to the one described in [19] was used to calculate the real part $f_1$. Finally the optical theorem [27] links the optical constants δ and β to the forward scattering cross sections $f_1$ and $f_2$ according to: $\delta(E) = \frac{N_A \rho}{A} \frac{r_0 \lambda^2}{2\pi} f_1(E)$ and $\beta(E) = \frac{N_A \rho}{A} \frac{r_0 \lambda^2}{2\pi} f_2(E)$. Here, $r_0$ is the classical electron radius, $N_A$ is the Avogadro constant, A is the atomic mass number in (mass/mol), ρ is the mass density in (mass/volume) and λ is the wavelength of the incident X-rays. The results are shown in figure 3 for the energy range close to the Mn $L_{2,3}$ edges.

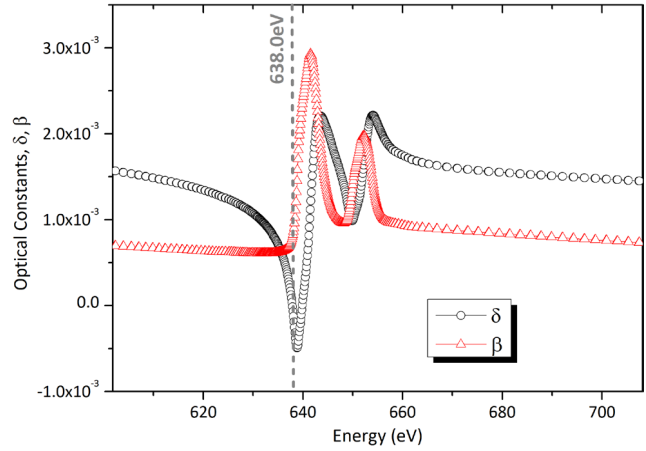

**Figure 3.** Optical constants for La$_{2/3}$Ca$_{1/3}$MnO$_3$ in the vicinity of the Mn L edges derived from measuring a XAS and applying a Kramers-Kronig transform (see details in the text). The indicated energy of E=638 eV was used for the XRMR investigation. As can be seen, the dispersion dominates at this energy while the imaginary part of the index of refraction is still small.

For the calculation of δ and β from f$_1$ and f$_2$ it was assumed that the stoichiometry of the LCMO layer is exactly La$_{2/3}$Ca$_{1/3}$MnO$_3$. The validity of this assumption will later be confirmed by the high quality of the structural reflectometry fits presented below. The indicated energy of E=638 eV is used for all subsequent XRMR measurements. This particular energy provides a large scattering cross section (δ is maximal) and nearly no resonant enhancement of the absorption. The optical constants of the YBCO layer and the STO substrate were directly retrieved from tabulated values [25]. This is justified since all resonant edges in these materials (Cu, Ba, and O) are far away from the energy where the XRMR is measured.

3.3 X-ray Resonant Reflectometry & Structural Characterisation

X-ray resonant reflectivity and XRMR are used to investigate the structural properties, i.e. layer thicknesses and roughness, of the bilayer system and the magnetic depth profile of the LCMO layer. The structural/chemical depth profile of the sample is obtained by measuring non-magnetic resonant reflectivity (XRR) curves, i.e. resonant reflectivity without XMCD contrast, and fitting a simulation to them. It is favourable to measure XRR and XRMR at the same energy because it allows using the results from XRR directly as input for the XRMR. The structural depth profile is determined by the layer thicknesses, the interface roughness and the chemical composition. The chemical composition was used as a fixed input parameter (nominal compositions), while the roughness and thicknesses of all layers and the substrate were set as free parameters of the fitting process.

We use the dedicated software tool "ReMagX" [16] for simulation and fitting of the reflectivity data. ReMagX uses a 4x4 matrix formalism [28] which allows simulating the reflectivity from an arbitrary magnetic layer stack. While standard reflectometry for optical isotropic media uses the Parratt formalism [29] for simulation this technique is not applicable to the resonant magnetic data presented here since since it neglects the optical anisotropic nature of the XMCD effect and therefore e. g. the itensity transfer from σ to π polarization and vice versa inside the stack. In contrast ReMagX calculates the full polarization dependent eigenfunctions for each layer and thus accounts for the anisotropy and any intensity transfer. Interface roughness and local variations of the magneto-optical properties and hence the magnetic moment is incorporated into the matrix description by slicing the sample [16,30]. The local variations of the magneto-optical constants can be introduced by Gaussian profiles with a variable position, FWHM and height. This either modifies an existing magnetic moment, i.e. magneto-optical contribution or it introduces a new, artificial magnetic signal. More details and also restrictions of the XRMR data processing and can be found e.g. in references [17-19,31].

Figure 4 shows the resonant non-magnetic reflectivity measured at T=110 K at an X-ray energy of $E_{h\nu} = 638$ eV using linear horizontally polarized light along with the best simulation result (red solid curve) obtained from fitting the data.

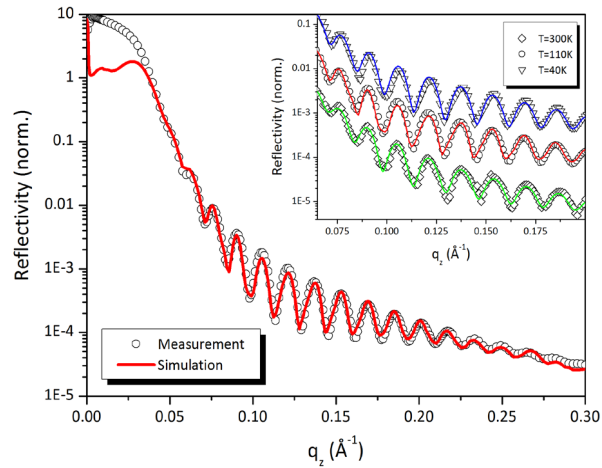

**Figure 4.** Resonant X-ray reflectivity curve without magnetic contrast measured at T=110 K using an X-ray energy of 638 eV and linear polarized light. The red curve is the best fit result. The inset shows a close up of the curve for $q_z$=0.0625 to 0.2 Å$^{-1}$ together with the results obtained at T=40 K and T=300 K. Note that the curves have been shifted vertically for better readability. No significant change of the non-magnetic reflectivity and therefore also the underlying chemical structure is found.

The simulation perfectly reproduces the measured reflectivity curve in the range from $q_z$=0.08 to 0.275 Å$^{-1}$ where the relevant, structure determined, features are located. The deviation for small angles is related to the incomplete illumination of the sample. The non-magnetic reflectivity was also measured for T=300 K and T=40 K to verify that the chemical structure does not change with temperature. A close-up of all three measured curves in the range from $q_z$=0.0625 to 0.2 Å$^{-1}$ is shown as inset in figure 4 together with the corresponding simulation results.

From the simulation shown in figure 4, the optical depth profile, which corresponds to the structural depth profile, is derived and the result is shown in figure 5.

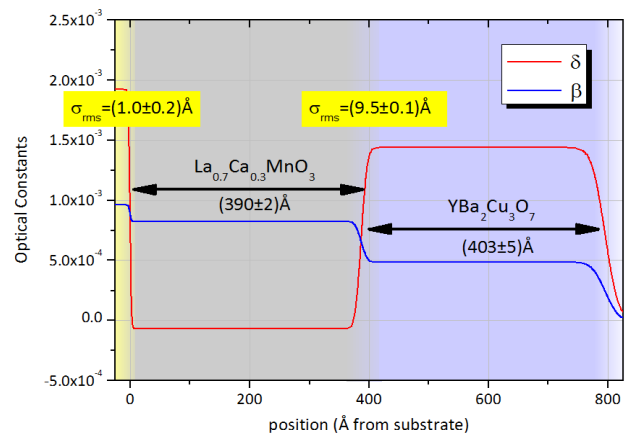

**Figure 5.** Optical, i.e. structural, depth profile of the sample measured at an X-ray energy of E=638 eV and a temperature of T=110 K.

A thickness of (390±2) Å and (403±5) Å is found for the LCMO and YBCO layer, respectively. Both interfaces are very smooth showing a RMS roughness of (9.5±0.1) Å for the LCMO/YBCO interface and only 1 Å for the STO/LCMO interface. The errors were derived from comparing the three temperatures measured.

## 3.4 X-ray Resonant Magnetic Reflectometry

A series of temperature dependent XRMR curves were measured for Θ ranging from Θ=0° to 27.75° (corresponding to $q_z$=0 to 0.3 Å$^{-1}$) in an applied external magnetic field of $H_{ext}$=130 mT parallel and antiparallel to the sample surface (in-plane geometry). X-rays with an energy of E=638 eV (Mn L$_3$ edge) with positive helicity were used and the external magnetic field was flipped at every angle value from positive to negative and the two specular reflected intensities were measured to obtain the circular dichroic reflectivity curves. Subsequently the measurement was repeated for negative helicity of the X-rays.

To extract the magnetic signal from the two dichroic curves, the asymmetry ratio between the two is calculated according to $A = (R^{\uparrow\uparrow} - R^{\uparrow\downarrow})/(R^{\uparrow\uparrow} + R^{\uparrow\downarrow})$. Here the first arrow indicates the helicity of the X-rays (↑= positive and ↓ = negative circular polarized light) and the second arrow indicates the direction of the external magnetic field (↑= positive and ↓ = negative) with respect to the direction of the incident X-rays.

First the magnetic asymmetry signal associated with Mn was measured at a temperature of T=110 K which is well above the superconducting transition temperature. Figure 6 shows the resulting asymmetry ratio together with the best simulation result obtained from fitting the data.

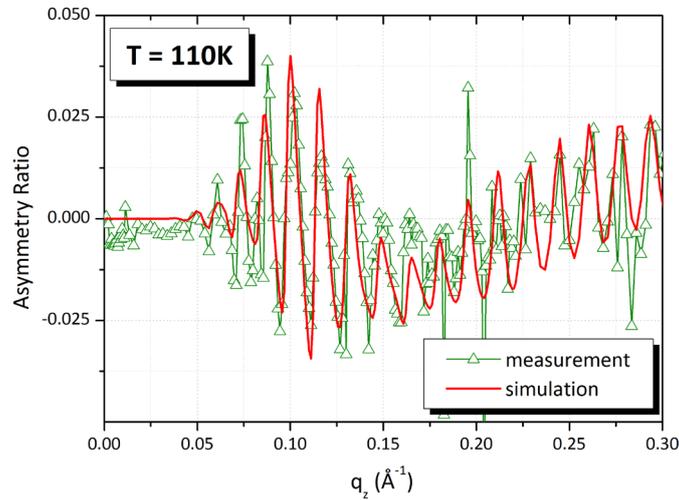

**Figure 6.** Element selective Mn asymmetry ratio between $R^{\uparrow\uparrow}$ and $R^{\uparrow\downarrow}$ measured at an energy of E=638 eV. The curve is the average of the positive and negative helicity measurements. The red solid curve represents the best fit result based on the magneto-optical profile shown in figure 7.

A small asymmetry ratio on the order of 2.5% is found originating from ferromagnetic Mn in the LCMO layer. This asymmetry ratio exhibits two major features: a short range oscillation which is mostly related to the YBCO thickness and a long range modulation which should originate from a smaller structure in the magneto-optical profile. Analyzing the short range oscillation yields $\Delta q_z$=0.0079 Å$^{-1}$ which corresponds to a thickness of 397 Å. This value corresponds, within the error, to the structural properties of the YBCO layer (compare figure 4). To obtain the magneto-optical profile connected with the asymmetry signal, the data has been fitted [16] and the fit result is also shown in the plot as a red solid curve. The fit reproduces the major features of the asymmetry very good. The fitting process consists of two steps: first magneto-optical constants for LCMO were derived from XMCD data of ferromagnetic LCMO and rescaled to account for the reduced magnetic moment of Mn in this sample (see Sec. III). In a second step the magneto-optical constants were used to fit the asymmetry signal while allowing local variations (enhancement or suppression) of the magneto-optical profile and hence the saturation magnetization. The result is shown as inset in figure 7.

The overall shape of the asymmetry ratio, especially the long range modulation can only be reproduced if a layer with reduced or completely suppressed magnetization close to the YBCO is inserted into the magneto-optical profile. Compared to the overall thickness of the LCMO film, this feature is quite narrow. Therefore, figure 7 shows a close-up of the interface region between LCMO and YBCO while the full magneto-optical profile is shown as inset in the same figure.

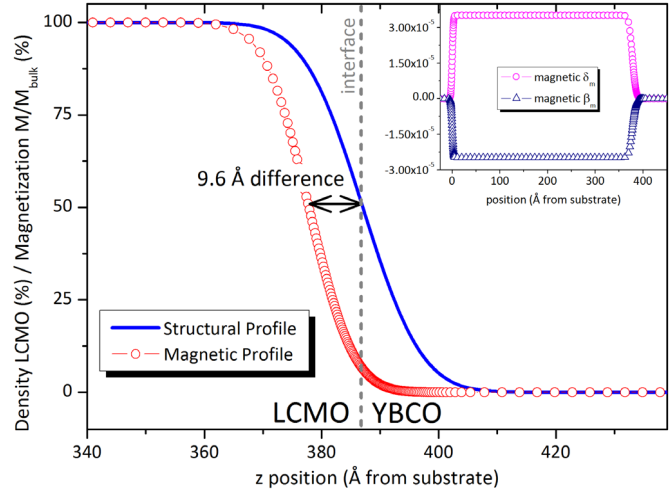

**Figure 7.** Magnified structural and magnetic profile close to the LCMO/YBCO interface. Note both graphs were normalized to the bulk value of the LCMO. The magnetic profile sets in roughly 10 Å beneath the structural transition. The magnetic moment close to the interface is strongly reduced. The inset to the upper right shows the absolute overall magneto-optical profile for the LCMO layer.

Here the Mn density (blue) and the magnetic signal (red open circles) are plotted in percent with respect to the bulk layer properties. The magnetic profile is shifted compared to the structural interface by 9.6 Å which means that the LCMO directly at the interface to the YBCO has a reduced or zero magnetic moment. Note, that the small asymmetry makes it difficult to derive the exact magnetization at the interface. Therefore we cannot exclude a small remaining moment for the LCMO directly at the interface; the magnetic signal in figure 7 can still refer to a ferromagnetic ordered state but with a strongly suppressed magnetization value.

The existence of a layer of suppressed LCMO magnetization at the interface to YBCO was already reported for some FM/SC superlattices [32,33]. Our results show that a suppressed layer already occurs at a single FM/SC interface and does not depend on the superconducting state of the YBCO in the bilayer system since it is already present at T=110 K. This latter observation is consistent with magnetization measurements of YBCO/LCMO bilayers which clearly show a YBCO thickness related decrease of the LCMO saturation moment already well above the superconducting transition (see figure 6 in reference [2]).

In contrast to the results presented here and in [32,33], a uniform magnetization of the LCMO layer, i.e. no suppression at the interface, was also found in superlattices [6,10]. A possible explanation for the contradictory results might be related to the roughness of the LCMO/YBCO interface. There is a striking coincidence of the interface roughness ($\sigma_{rms} = 9.5$ Å) as revealed by X-ray reflectometry and the shift of the structural and magnetic profile (see figure 7). This interface roughness – substantially larger than in the YBCO/LCMO superlattices [6,10] – could be accounted for the suppression of the magnetization at the interface. This interpretation is further stressed by recent high-resolution transmission electron microscopy (HRTEM) results of a YBCO/LCMO/YBCO trilayer revealed that the interface properties and also the roughness strongly depend on whether YBCO is grown on top of LCMO or vice versa [34]. In our case we find a larger interface roughness for YBCO grown on LCMO with reduced interface magnetization in the LCMO. The second explanation is related to the thickness of the YBCO layers. As shown in figure 6 of Ref. [2], the magnetic moment in LCMO decreases for larger YBCO thickness, revealing suppression of the FM. While the superlattices in Ref. [6,10] were prepared for thin YBCO (100 Å) the samples investigated here have four times higher thickness, consistent to the observation of the reduced FM.

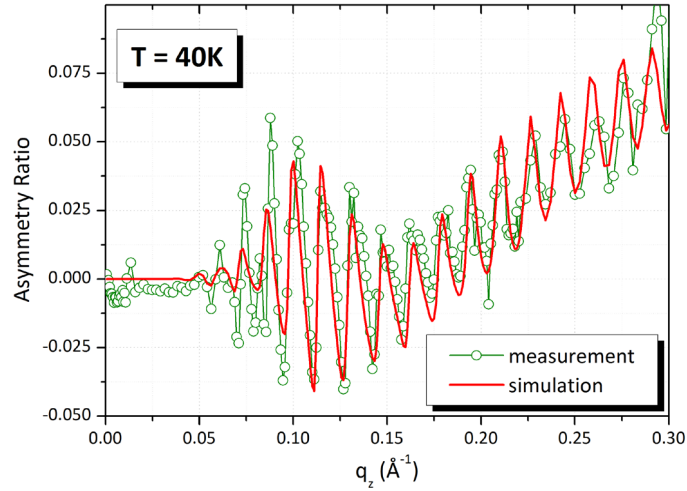

**Figure 8.** Magnetic asymmetry ratio measured at T=40 K and an X-ray energy of 638 eV.

Figure 8 depicts the analogue measurement to figure 6, now at a temperature of T=40 K, much below the superconducting transition of the YBCO layer. Similar to the high temperature data, we can simulate the asymmetry nearly perfectly by our algorithm.

In addition to that, we find a striking difference between the results at T=110 K and T=40 K, respectively. The asymmetry curve at T=40 K exhibits a clear increase at $q_z$ values of $q_z$=0.2 Å$^{-1}$ and above. A wide maximum seems to occur for high momentum transfer. Since the observed reflectometry oscillations refer to inverse length scales as in reciprocal space, a wide maximum refers to a narrow structure in the magnetization profile. Moreover, the fact that it occurs at large $q_z$ values directly indicates that this feature has to be buried in a large depth under the surface.

Our algorithm again allows the determination of the corresponding magnetic profile which is depicted in figure 9. At a temperature of T=40 K a new feature of the magnetization profile across the LCMO layer is found. A sharp peak where the local magnetization is roughly doubled is found close to the interface to the substrate. Obviously there is an interaction between the LCMO film and the STO substrate a low temperatures leading to an additional magnetic effect.

The presence of a magnetization increase close to the substrate in a YBCO/LCMO superlattice was already reported from neutron scattering [35]. We confirm this finding and show in the following an analysis of the temperature dependence of the effect providing deeper insights into its origin.

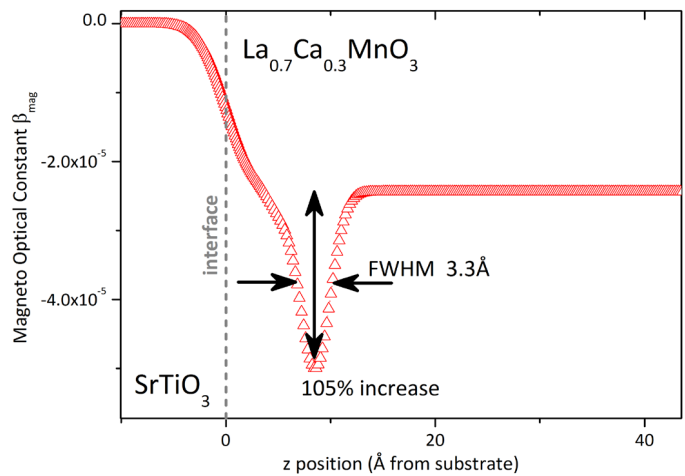

**Figure 9.** MO profile for the absorptive part of the index of refraction in the vicinity of the LCMO/STO interface obtained for the T=40 K asymmetry. The MO constants are both increased by more than 100% which indicates an increase of the Manganese magnetic moment per atom to $\mathbf{1.44\mu_B}$ with respect to the SQUID derived bulk value.

To discuss this finding systematically, we performed the analogue analysis to figure 6 and figure 8 at additional temperatures. For a quantitative statement, the asymmetry curves at T=40 K, T=65 K, T=80 K, and T=110 K have been fitted and the result is plotted in a new graph which is shown in figure 10.

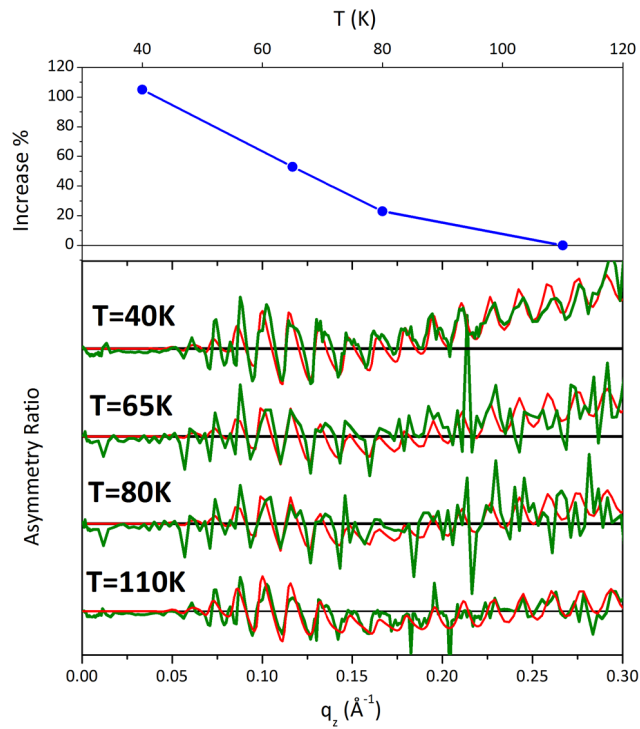

**Figure 10.** Temperature dependence of the increased Mn magnetic moment close to the STO interface. The lower part shows all measured asymmetry curves along with the corresponding fit results. The upper section shows the percent increase of the Mn magnetization at the STO interface (see figure 9) derived from the simulation curves.

All four temperatures can be fitted using the same general shape of the magneto-optical profile close to the STO interface (compare figure 9). The increased noise level of the T=65 K and T=80 K measurements is related to a reduced integration time for these measurements. Nevertheless the general trend is clearly visible from the curves and can be fitted successfully. We clearly see that the effect continuously decreases with increasing temperature and seems to vanish around T=100K.

The observed temperature-dependence of the LCMO/substrate interaction is tentatively considered to be related to the structural properties of STO single-crystals at low temperatures. Below T=105K a structural phase transition from cubic to tetragonal occurs which in our case directly leads to strain at the LCMO/STO interface [36,37]. Since the temperature dependence of the additional magnetization peak looks similar to the structural data in the same temperature range (see e.g. in [37]) it is reasonable to imply a direct correlation of the magnetization increase and the tetragonal distortion of the STO.

This result is contradictory to other investigations about the effect of compressive strain on the magnetic properties of LCMO which show that the magnetization is usually decreased in such films [38-41]. However these investigations used samples with a static strain e.g. by using different substrates or by varying the layer thickness. Moreover they were obtained by using standard magnetometry which measures the whole film and not a local variation of it. The investigation presented here is very different in two important points: the XRMR technique detects even small localized changes of the magnetization profile. In the present case the observed increase at the LCMO/STO interface is less than 1% (for T=40 K) of the total LCMO magnetic moment. Such a small signal is hard to detect with standard, bulk sensitive magnetometry. The second important difference is that the increase of the magnetization observed in the present work is reversible. The measurement sequence was 40 K → 110 K → 64 K → 80 K and the effect is visible at T=40 K, gone at T=110 K and comes back when cooling again to 64 K (see figure 10). This implies that the strain which is responsible for the effect is reversible. Therefore it might not be possible to directly compare the current results with investigations on growth induced strain.

Instead we propose a scenario where the growth induced strain from the substrate [41] and strain caused by the presence of the YBCO layer [2] create a thickness dependent state of stress in the LCMO layer which is responsible for the reduced magnetic moment in the bulk observed here. The tetragonal distortion of the STO below T=105 K disturbs this balance and allows a relaxation of the first few monolayers of the LCMO which causes a local increase of the saturation moment. This scenario would also explain why the effect was not yet observed for pure LCOM layers (see e.g. in [42]).

## 4. Conclusion

In conclusion, we have found that the magnetization profile across ferromagnetic LCMO layers in bilayers of LCMO/YBCO on STO substrates is strongly affected by the interfaces to the adjacent materials. At all temperatures we find a suppression of the LCMO magnetization close to the YBCO layer on a length scale of 10Å, the transition to superconductivity seems not to influence this finding. A second effect occurs at the LCMO/STO interface at low temperatures. Owing to the structural phase transition in STO single crystals at T=105 K a peak-like magnetization increase develops inside the LCMO layer close to the substrate interface which has its origin in strain arising from the STO substrate.

## Acknowledgement

The authors are indebted to B. Keimer, E. Benckiser, and V. Hinkov for helpful discussions. Support by B. Zada, W. Mahler and the whole BESSY II team during the beamtime is gratefully acknowledged.